# Temperature profile and double images in the inferior mirage

Nabil W Wakid[1*]

[1] Department of Chemistry, Physical Science Section, Middle Tennessee State University, Tennessee 37132

[*] nabil.wakid@mtsu.edu

Desert mirages were simulated in the laboratory by heating a flat surface of sand. This showed that a boundary layer of air, of only a few millimeters immediately covering a heated surface has nearly the same thickness over a wide range of temperatures. It consists of a region producing an inverted image, and another producing an erect image depending on the relative distances between observer, heated surface and object. Measurement of distances and heights show angles of incidence greater than angles of reflection. Air refractive index measurements agree with involvement of the critical angle in ray bending. An analogy with density gradient sucrose solutions also shows double images, inverted and erect. It is assumed that a similar situation exists outdoors**.** Specular reflection over rough surfaces can be described as a mirage, but it is seen at a wider range of viewing angles, and it does not require a temperature gradient. It is always present, but it can be obliterated by a higher intensity of diffuse reflection.

## Introduction

Mirages that are seen in the desert are more commonly observed as what appears to be a puddle of water on the road while driving on a sunny day. What is seen in this apparent puddle is a virtual inverted image of a distant object. It is called "inferior", because it appears reflected below the real object. Most descriptions of this phenomenon ascribe it to refraction through less dense layers of air over a hot surface [1—6]. See also an extensive review by Young [7]. The earliest attempts at explaining mirages explored the possibility of parabolic and hyperbolic ray paths through refractive index gradients of warm air [1]. More complex ray paths were later considered [3 — 7] including phenomena where double images were observed [8 – 10]. However, an apparently overlooked paper in 1959 by Raman and Pancharatnam [11] explained that according to Snell's law of refraction, reflection must occur when $n_1 \cos\phi > n_2$, where $n_1$ is the refractive index of cooler air, $n_2$ the refractive index of heated air and $\phi$ the glancing angle of incidence. In other words, reflection occurs when $n_1 \sin\theta > n_2 \sin 90º$, where $\theta$ is the angle of incidence with respect to the normal. The incident ray is refracted away from the normal as it enters regions of lower refractive index, changes direction when $n_1 \sin\theta = n_2 \sin 90º$ (i.e. when $\theta$ is the critical angle) and subsequently is refracted toward the normal as it continues through increasing refractive index. Thus, the ray path is continuously curved throughout the gradient. This was beautifully demonstrated by Greenler [12] in a tank containing a gradient refractive index of salt in water (Plate VII) without any reference to the work of Raman and Pancharatnam.

Here, simulation of the desert mirage in the laboratory indicates that the temperature gradient responsible for reflection is confined within a thin boundary layer immediately over the heated surface. The thickness of this layer is shown to be unchanged over a wide range of temperatures. Ray tracing shows that the appearance of double images depends on the relative distances between observer, heated surface and object. An analogy with gradient density sucrose solutions also shows why double images occur.

Others explained a mirage as merely due to specular reflection, and that it is independent of weather conditions [13—16]: light striking a rough surface at a grazing angle (i.e. at a large angle of incidence) is not reflected diffusely but exhibits mirror-like reflection. So, this work also includes a comparison with specular reflection over rough surfaces.

# Materials and methods
## Mirage simulation

Mirage simulation was carried out with an electric food warmer covered with a layer of sand approximately 54 cm long, 21 cm wide and 0.3 cm thick (Fig 1). Temperatures were taken with a thermometer placed on the sand. The objects viewed were drawings of colored circles and oblique lines. All photographs were made by the author with a Canon Powershot GLPH150IS having 10 x optical zoom.

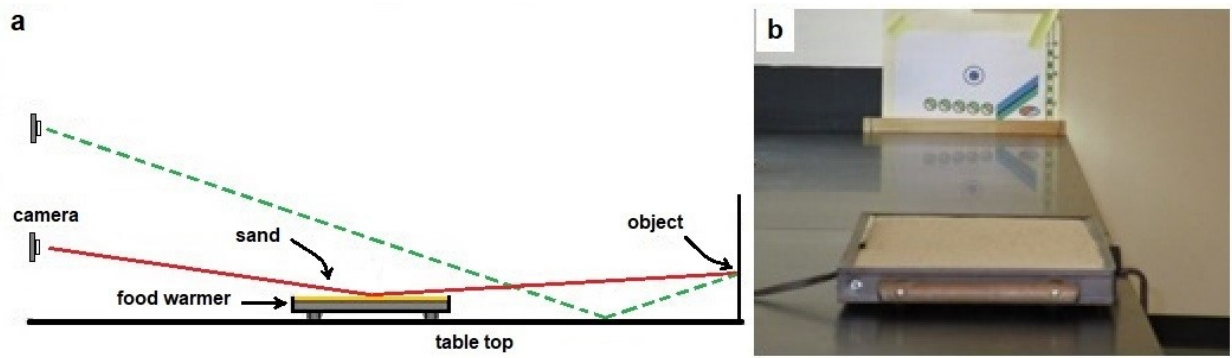

**Fig 1. Set up for simulation of a mirage in the laboratory.** (a) Diagram showing ray tracing in red for reflection over the heated sand layer. The dashed line shows specular reflection over the table top. Distance between camera and object was 570 cm. Mirages were observed by varying the height of the camera and by moving the food warmer between camera and object. (b) Photograph of the food warmer in the foreground with layer of sand and thermometer. The sand surface was 3.8 cm over the table top. Note the specular reflection over the black epoxy resin table top. Object height, center of fourth circle from left, was 4.0 cm; camera height 30 cm; angle of incidence 86.6° (Eq 1).

## Temperature measurements

Outdoor ambient air temperature was taken with a regular laboratory thermometer having a precision of 0.1 °C. Air temperature adjacent to the ground was measured by placing the thermometer on the pavement at the shoulder of the road with the assumption that it is representative of the entire sunlit area. Indoors, a temperature profile above the heated sand layer was obtained with thermometers taped onto a meter stick held vertically over the sand layer with utility clamps. Thermometer heights were 0-, 1-, 3-, 13-, 55-, and 89 cm above the sand surface. Readings were taken as the sand temperature increased.

## Distance measurement

A measuring tape graduated in millimeters was used for indoor distances. Outdoor distances were measured with a mechanical odometer, fitted at the end of a cane, equipped with a wheel of known circumference and a digital readout in feet and inches. The angle of incidence for specular reflection was measured geometrically from its tangent:

$$\tan \theta_i = \frac{b}{H} = \frac{d-b}{h} \tag{1}$$

where $H$ is the height of the object; $h$, the height of the camera; $d$, the distance between camera and object, and $b$ the distance between the object and the vertex of the incident angle. The value of $b$ was calculated from the measurements of $d$, $h$ and $H$ (Fig 2).

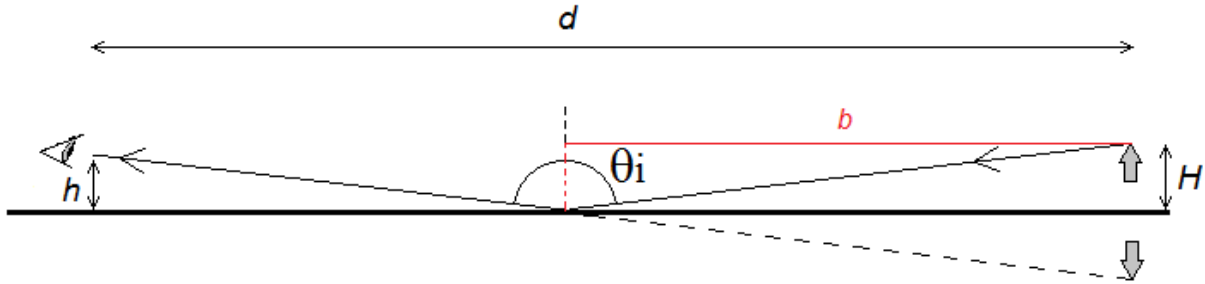

**Fig 2**. Measuring the angle of incidence for specular reflection.

## Air refractive index

The refractive index of air was determined by using the equation of Stone and Zimmerman [17]:

$$n = 1 + \frac{7.86 \times 10^{-4} p}{273+t} - 1.5 \times 10^{-11} RH(t^2 + 160) \tag{2}$$

where $p$ = atmospheric pressure in kilopascals, $t$ = temperature in degrees Celsius, and $RH$ = percent relative humidity. Atmospheric pressure was measured with a calibrated aneroid barometer. Ambient air temperature was taken with a regular laboratory thermometer having a precision of 0.1 °C. Air temperature adjacent to the ground was measured by placing the thermometer on the pavement at the shoulder of the road with the assumption that it is representative of the entire sunlit area. A wet bulb thermometer was used for measuring the relative humidity, $RH$, of ambient air. Relative humidity of air adjacent to the ground, $RH_G$, was determined using the Clausius-Clapeyron equation [18]:

$$RH_G = 100 \times \frac{EXP\left(-\left(\frac{40700}{8.3145}\right)\left(\frac{1}{t} - \frac{1}{373}\right)\right)}{EXP\left(-\left(\frac{40700}{8.3145}\right)\left(\frac{1}{T} - \frac{1}{373}\right)\right)} \tag{3}$$

in which $t$ is the dew point temperature, and $T$ the air temperature over the heated surface. The dew point was obtained from the relative humidity of ambient air, $RH$, by using the equation of Lawrence [19] in which $T_A$ is the ambient temperature:

$$t \approx T_A - \frac{(100-RH)}{5} \tag{4}$$

## Height of the vanishing line

The height where the oblique line changes direction, also called the vanishing line [5], was measured near the center of the warmer using printed photographs as shown in Fig 3.

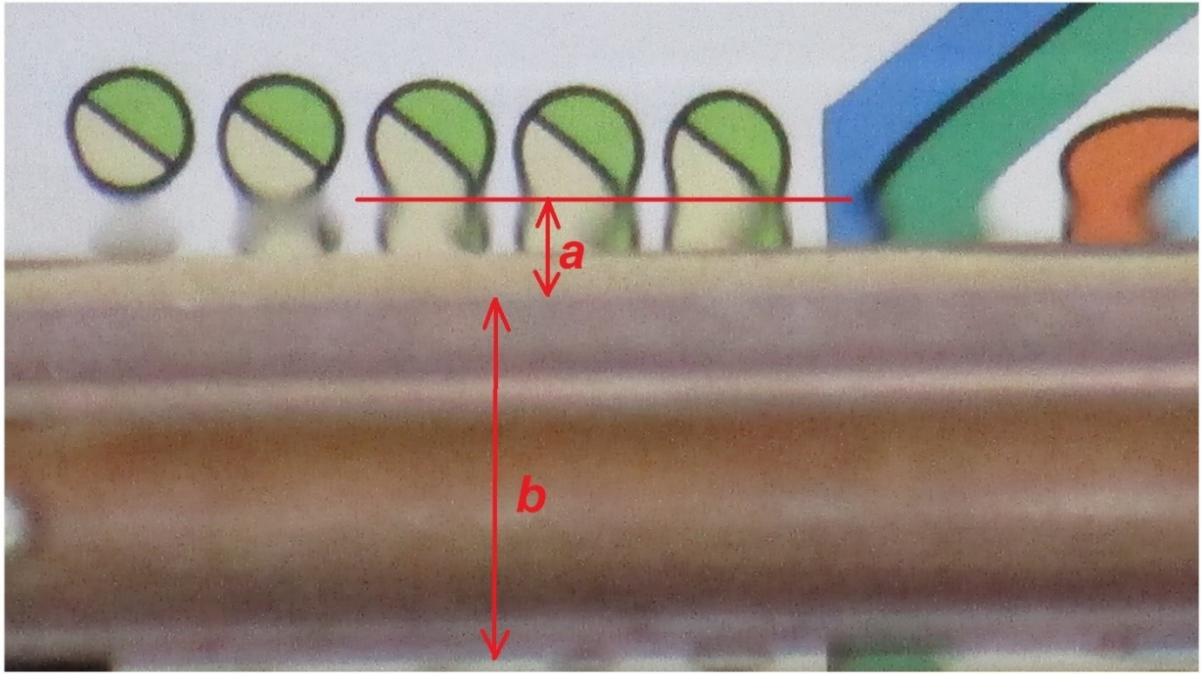

**Fig 3. Height of the vanishing line** (red horizontal line) was determined by measuring the lengths of '**a**' and '**b**' on photographic prints. Height = 30 x a/b, where '**a**' is the height of the vanishing line with respect to a fixed reference – the edge of the warmer, '**b**' is the measured thickness of the food warmer and 30 is its actual thickness in millimeters. In this photograph the middle of the food warmer was 310 cm from the camera which was 5.0 cm above table top. Sand temperature was 56 ºC.

## Sucrose gradient

A tank made of 3 mm-thick sheets of Perspex was used for gradient solutions of sucrose. Internal dimensions were 15.2 x 7.0 x 12.5 cm. The sucrose solution at the bottom was layered gradually with distilled water using a Pasteur pipette.

# Results and discussion

## Outdoors

Fig 4 shows two stretches of road on sunny and cloudy days. With an observer height of 1 m mirages were only seen on sunny days with temperature differences above 7º C. No mirage was seen when temperature differences were low. This indicates that relatively high temperature differences are needed to form a boundary layer of warm air over the road surface. See **Temperature profile** below. Also, from the Snell equation it can be appreciated that as the difference in refractive index (or difference in temperature) decreases, the value of the critical angle approaches 90º, and it then becomes more difficult to see a mirage. So, mirages are more readily seen at lower critical angle values. That no mirage was observed on cloudy days indicates that the reflection seen on sunny days is due to refraction caused by increased heating of air adjacent to road surfaces. If the mirage seen on sunny days were due to specular reflection, it would also be seen on cloudy days when differences between ground and ambient air temperatures are low.

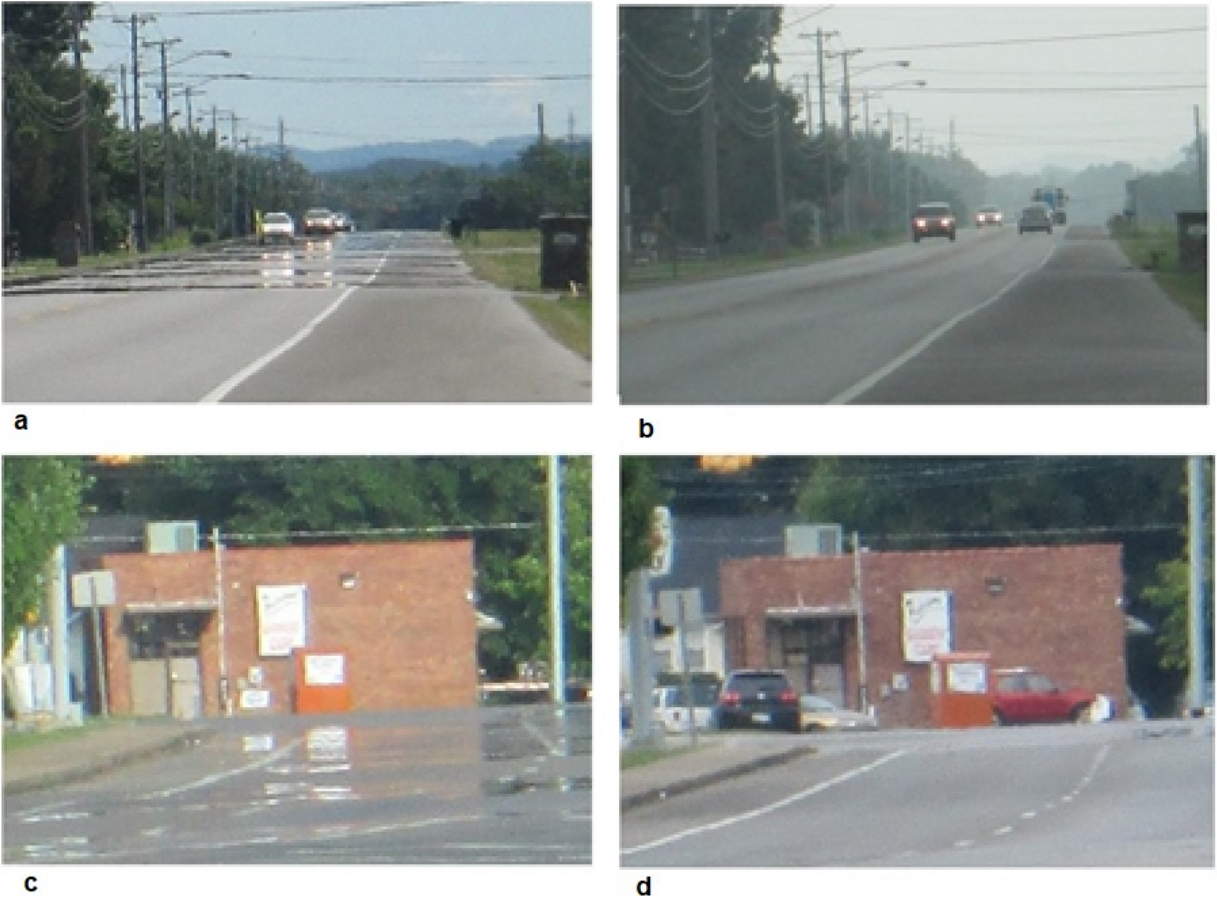

**Fig 4. Two road views** on sunny (left) and cloudy days. Temperature differences between road and ambient airs were 13 °C (a), 3 °C (b), 10 °C (c) and 6 °C (d).

Reflections in Fig. 4 appear as stripes on the surface of the road. This phenomenon has been described in detail by van der Werf [9] who attributed these irregularities to reflections from undulations in the road surface. To avoid this effect, simulation of a mirage in the laboratory was undertaken where it was possible to use an even and flat horizontal surface of sand for observing an inverted image.

## Indoors

A simulation of mirages in the laboratory has already been demonstrated by Fabry et al. [8], by Greenler [12], and by Vollmer and Tammer [20, 21]. Here, a setup is presented that can demonstrate both, refraction and specular reflection (Fig 1). The object, a drawing of circles and oblique lines, was viewed over a heated surface of sand. At increasing temperatures, they appear to be reflected over the sand layer (Fig 5). Note that the bottom of the 3$^{rd}$, 4$^{th}$, and 5$^{th}$ circles is no longer visible, nor is the 3-cm mark on the scale to the right. There is an imaginary horizontal line below which the objects cannot be seen. To quote Greenler [5], "the inverted image appears to be a reflection at this vanishing line." Bravais [1] called it "la ligne de partage" (the dividing line). Figs 5c and 5d show that the height of this line did not change with increasing temperature. Measurements made on 10 additional photographs of laboratory mirages show that the height of the vanishing line does not change appreciably over a wide range of temperatures (Fig 6a). Judging from Fig. 3 and Fig 6a, the height of the line over the sand would be less than 7.7 mm, perhaps half this value. This indicates that after entering the refractive index gradient, the light ray changes direction at a limited height above ground surface.

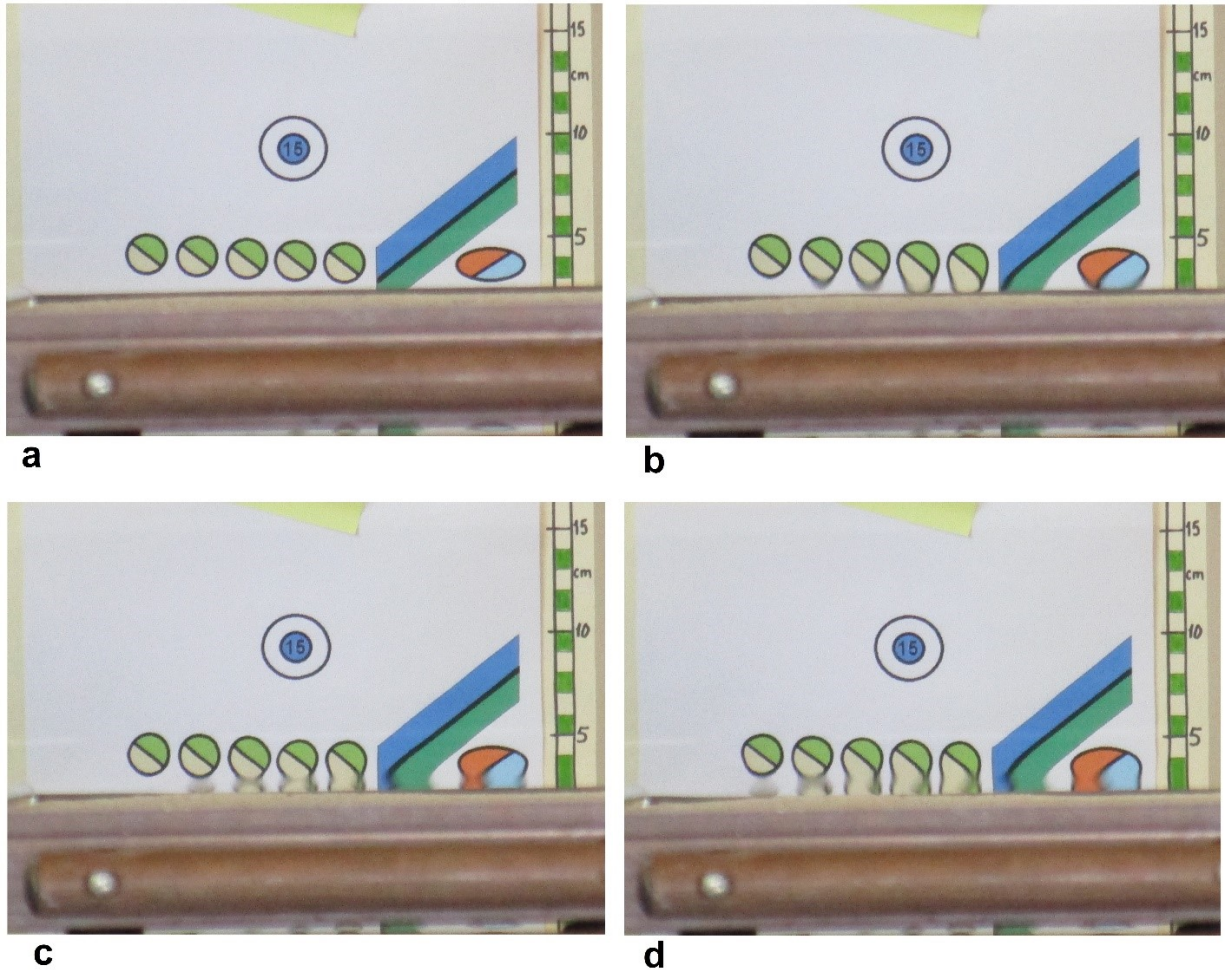

**Fig 5**. **Mirage simulation in the laboratory**. See Fig 1 for setup. (a) View at room temperature, 22 °C, with camera lens at 5.9 cm and sand 3.8 cm from table top. (b) Distortion due to refraction through warm layer of air over the heated sand at 31 °C. (c and d) Apparent reflection of circles and oblique lines over the sand layer at 50- and 61 °C respectively. Distance from camera to middle of heated surface was 320 cm and from camera to object 570 cm.

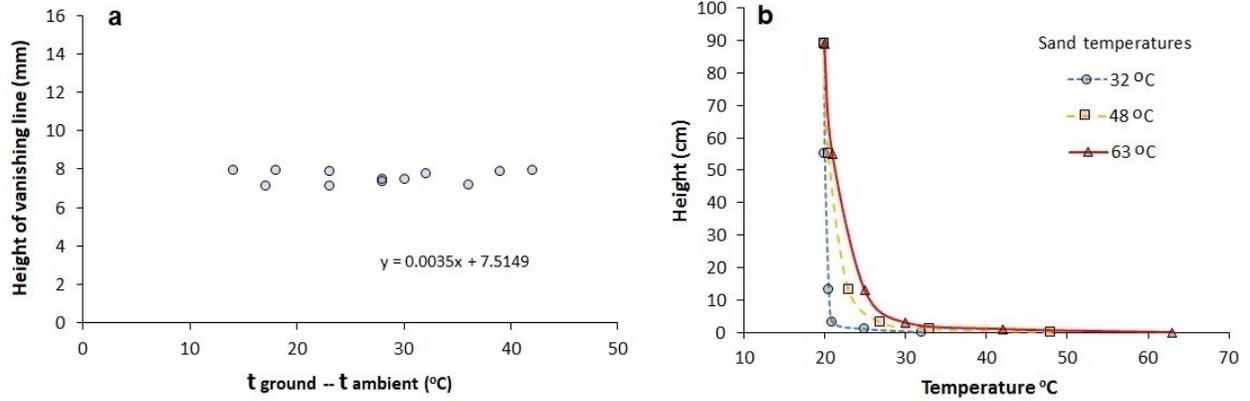

**Fig 6. Effect of temperature difference on the height of the vanishing line (a).** Average height was 7.7 mm. Camera height was 5.9 cm and its distance to the middle of the sand was 320 cm. **Temperature profile over sand layer (b).** The ambient temperature was 20 °C.

**Temperature profile**

A temperature profile over the heated surface (Fig 6b) shows that the greatest temperature change occurs within the first one to two centimeters. Detail of air temperatures below 1 cm is not feasible (even with thin thermocouples) because at close proximity to the surface, heat transfer by radiation becomes significant. What is changing therefore, is the refractive index of air within the confined height of a boundary layer only a few millimeters thick over the heated surface. This layer builds up starting with a temperature difference $\Delta T \geq 0$ °C and reaches a plateau at or before $\Delta T = 14$ °C (Fig 6a). Within this thin layer the pressure gradient is essentially nil, and heat transfer occurs mostly by conduction which approximates that in a solid dielectric slab [22]. As air temperature rises, molecular velocity increases resulting in more frequent collisions (increased viscosity) which favors heat transfer by conduction rather than convection.

In simple terms, the heat flow, Q, from an area, A, per unit time, is directly proportional to the temperature difference and inversely proportional to the thickness of the layer. It can be expressed as follows [22]:

$$Q = k \frac{A(t_1 - t_2)}{L} \qquad (5)$$

where $A$ is the area of the heated surface in m$^2$ and $L$ the thickness of the layer in m. The proportionality constant, $k$, called the thermal conductivity (measured in watts per meter Kelvin) is a transport property of the boundary layer. Here, the value of $L$ can be considered constant because it is invariable over a wide range of temperatures (Fig 6a), and the area is always the same; so, the heat flow, Q, is mainly dependent on the difference in temperature. Thus, heat transfer is in a steady state in the vertical dimension ($z$): heat gained from the sand is lost above this layer through natural convection. Changes in temperature difference affect the steepness of the refractive index gradient, $\delta n/\delta z$, not the thickness of the layer. The value of $k$ at 300K is listed as 0.026 W/m K [22]. Eventually, extremes in temperature or pressure can influence the physical properties of the medium and hence the value of $k$, which is therefore not a true constant.

It is only above this boundary layer that buoyancy forces due to the earth's gravity can occur by natural convection. This would be the region of turbulence above the vanishing line, approximately between 1- and 90 cm (Fig 6b). The stresses caused by shear flow of incoming cooler air can also produce turbulence in the boundary layer. Although there is also a temperature gradient above the vanishing line, it is not responsible for image reflection. There are reports of temperature profiles over mirages [23, 24], but these do not include measurements in close vicinity to the heated surface.

**Double images**

Double imaging had been observed by Bravais [1] and Tränkle [10] outdoors, by Wollaston [25] against a heated iron rod, by Minnaert [2] against a sunlit wall and by Fabri *et al.* [8] against a vertical heated metal plate. Tränkle attributed double imaging to an "unusual temperature profile" between observer and object.

Here, double imaging was observed by changing the position of the food warmer between camera and object. At long range, only inverted reflection is seen (Fig 7a), while at closer range to the camera double images appear: an inverted image and an upright image below it, immediately next to the heated surface (Fig 7c). All reflections disappeared when the food warmer was closer than 150 cm to the object. This shows that the appearance of double images depends on the viewer's perspective and that double images would be an integral part of the inferior mirage.

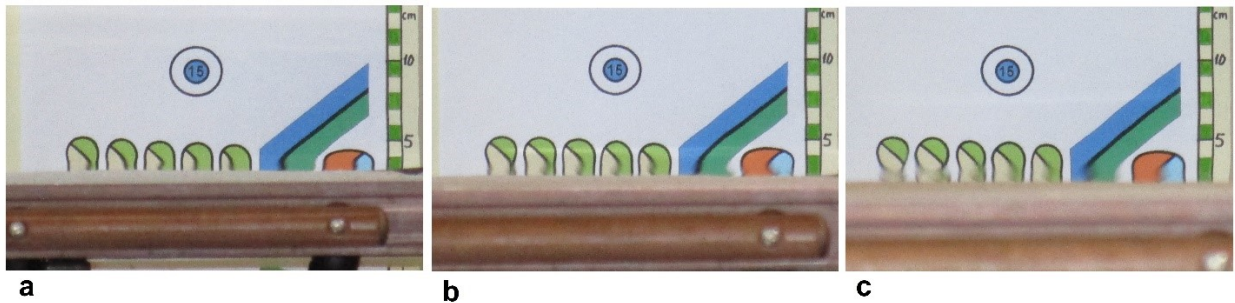

**Fig 7. Effect of distance to heated surface.** Distance between camera and middle of food warmer was 390 cm (a), 340 cm (b) and 270 cm (c). Total distance between camera and object was 570 cm; camera height was 5.7 cm and centre of fourth circle from left 4.0 cm above table top. Room temperature was 22 °C and sand temperatures were 77 °C (a), 75 °C (b) and 74 °C (c).

To understand double imaging an approximation of ray paths was attempted using as object an oblique line graduated in centimeter segments of different colors. In Fig 8a reflections (where the oblique line changes direction) appear at approximately 4.0 cm above the table top at long range, but at short range reflections appear at approximately 3.6 cm (Fig 8b). To trace a light ray, let us consider for example the situation in Fig 8b. The angle of reflection would be 89.62° (arctan 270 cm/(5.6-3.8) cm)). However, the tangent of the angle of incidence would be negative, i.e., 300 cm/(3.6-3.8 cm). This would make an angle of incidence greater than 90° (arctan -1500 +180°). The asymmetry between angles of incidence and reflection is shown in Fig 10a. (See S3 Table.) Changing the camera height also showed an angle of incidence greater than the angle of reflection even with the food warmer halfway between the camera and object (Figs 9, 10b. (See S4 Table.) This asymmetry may be explained if the light ray is entering at a point where the refractive index is lower than at the point of exit.

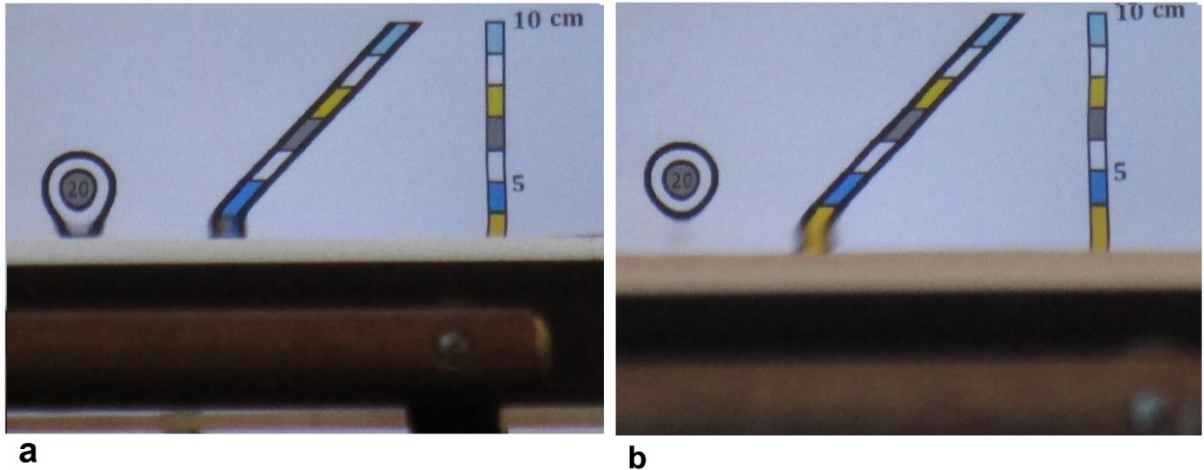

**Fig 8. Effect of distance using a graduated oblique line.** Distance between camera and middle of food warmer was 390 cm (a) and 270 cm (b). Total distance between camera and object was 570 cm. The sand layer was 3.8 cm from the table top and camera height, measured from the center of the lens, was 5.6 cm. Sand temperatures were at 71 °C. These observations are included in Fig 10a.

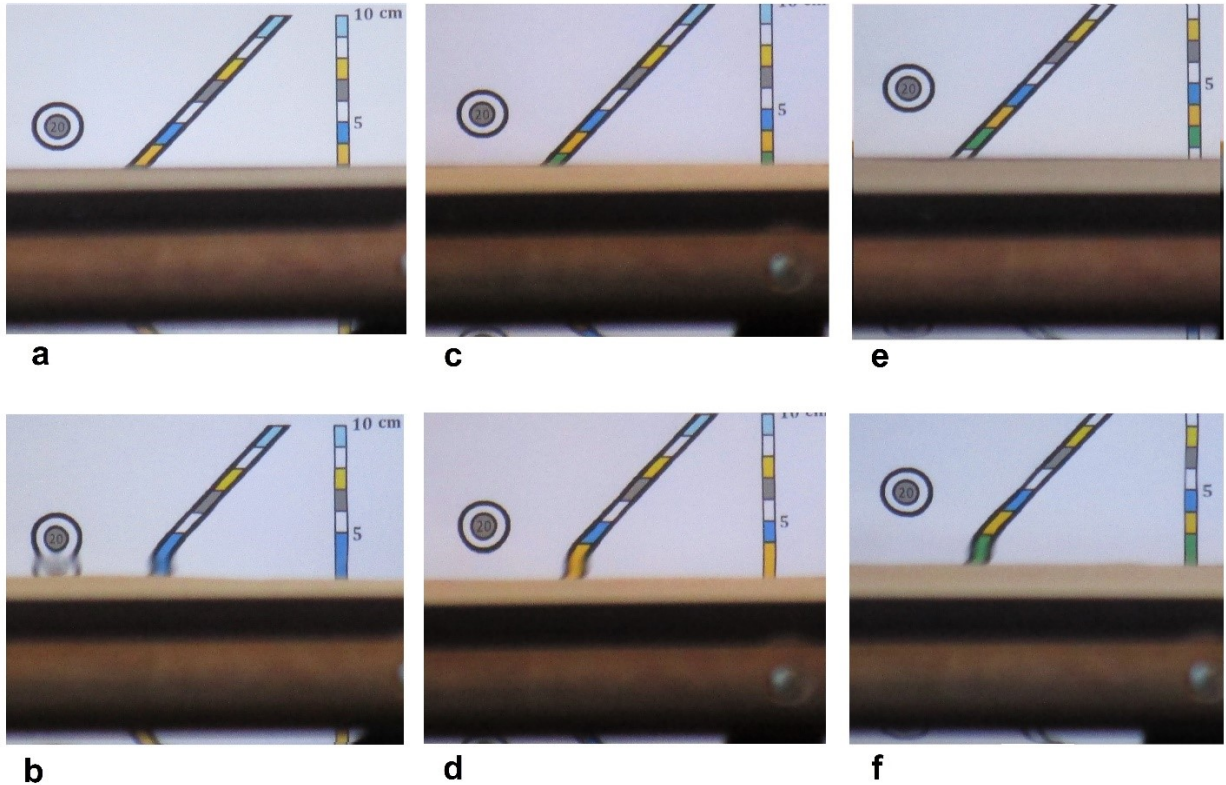

**Fig 9. View from different observer heights**. Camera height above the table top, measured from the center of the lens, was 5.1 cm (a) (b), 5.8 cm (c) (d) and 6.8 cm (e) (f). The top row was at room temperature, 22 °C. Sand temperatures were 69.5 °C (b), 66.5 °C (d), and 71 °C (f). Distance between camera and object was 570 cm with the center of the food warmer exactly halfway between the two. The sand in the food warmer was 3.8 cm above the table top. These observations are included in Fig 10b).

Fig 11(a) shows approximate ray paths in a refractive index gradient over a heated surface. An object, ABCD, appears reflected to an observer at point O. Part AB of the object is seen inverted and part CD erect and slightly compressed. This is representative of what is shown in Figs 7b, 7c, 8b, 9b, 9d and 9f. In Fig 11b the heated surface is closer to the object and only the inverted part AB is seen by the observer at O (as in Figs 7a and 8a).

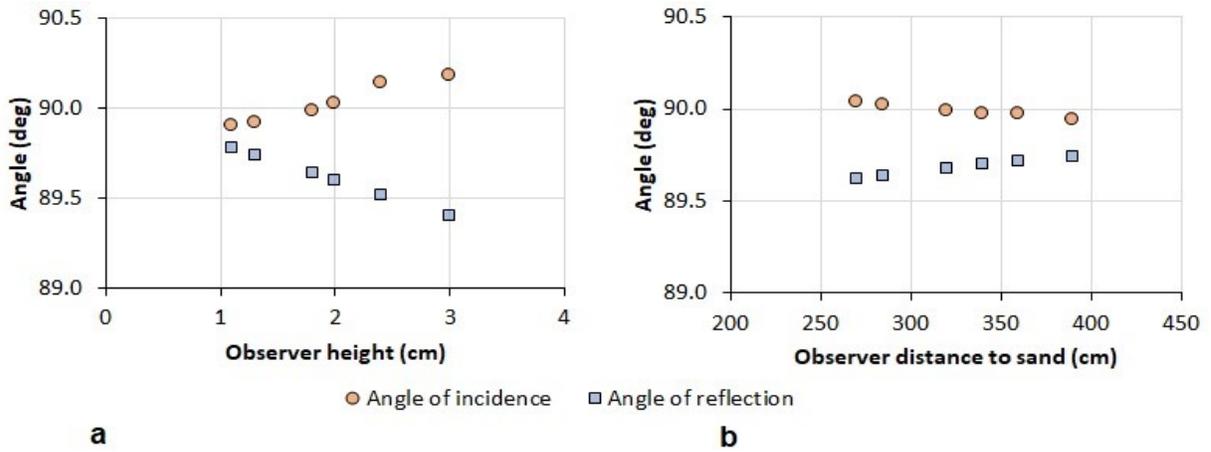

**Fig 10. Effect of observer distance (a) and observer height (b)** on the angles of incidence and reflection. Sand temperatures ranged between 64- and 71 °C.

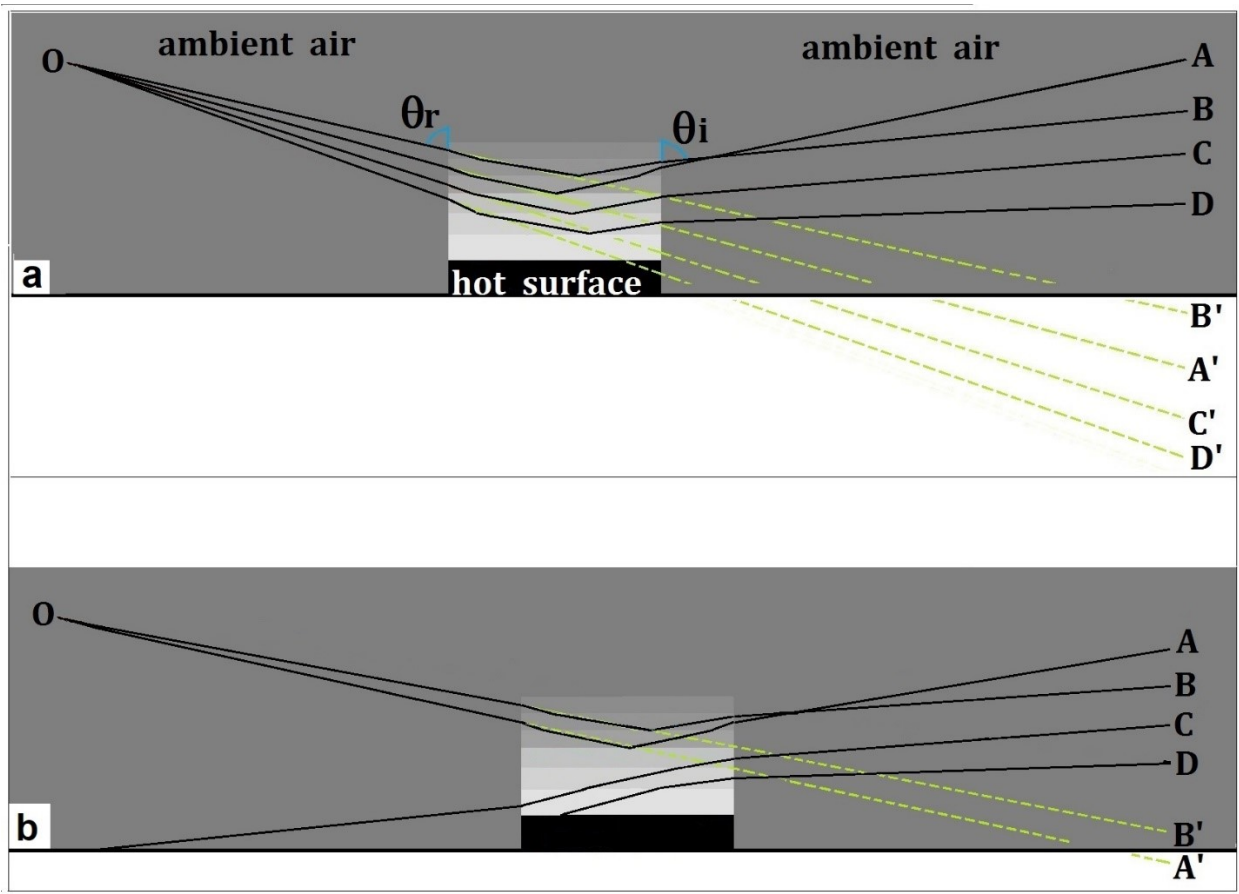

**Fig 11. Ray diagram over a heated surface depicting inverted and non-inverted imaging.** The vertical scale is exaggerated to show the gradient refractive index in layers (increasing darkness with increasing refractive index) but which is actually continuous. In (a) points A and B appear inverted to the observer at O because their rays cross. The angle of incidence, $\theta i$, is greater than the angle of reflection, $\theta r$. However, within the gradient, light rays are changing direction at the precise level where their angle of incidence exceeds the critical angle. Points C and D appear erect: their rays do not cross. In (b) the heated surface is closer to the object and only the inverted image appears.

It may not be practical to measure the refractive index of air at different levels in a thin gradient, but it is possible to calculate an average value for the critical angle from measurements of the refractive index of ambient air, $n_1$, and air adjacent to the ground, $n_2$, using Eq (2). In Fig 10 the average critical angle was 89.5°, a value fairly close to that of the incident angles considering that measurement of the refractive index does not involve measurement of length. (See S5 Table.) These results agree with the involvement of the critical angle in ray bending.

Of course, these observations rule out specular reflection, because if the sand layer were a flat mirror, the angles of incidence and reflection would coincide. It is difficult to imagine angles of incidence greater than 90°, because outdoors this would mean that the object is below ground level. This would be possible, nonetheless, when observing against a heated wall (See Minnaert [2], Plate Va).

**Sucrose gradient**

The gradient refractive index of air over a hot surface has been compared to that of gradient aqueous solutions [12, 21, 25, 26], but with gradients reversed: the high refractive index in solutions being at the bottom. As far back as the 19$^{th}$ century, Wollaston [25] experimented with gradient sugar solutions in an attempt to explain double imaging. Here, Wollaston's experiment was repeated, this time with an oblique line graduated in centimeter segments of different colors (Fig 12).

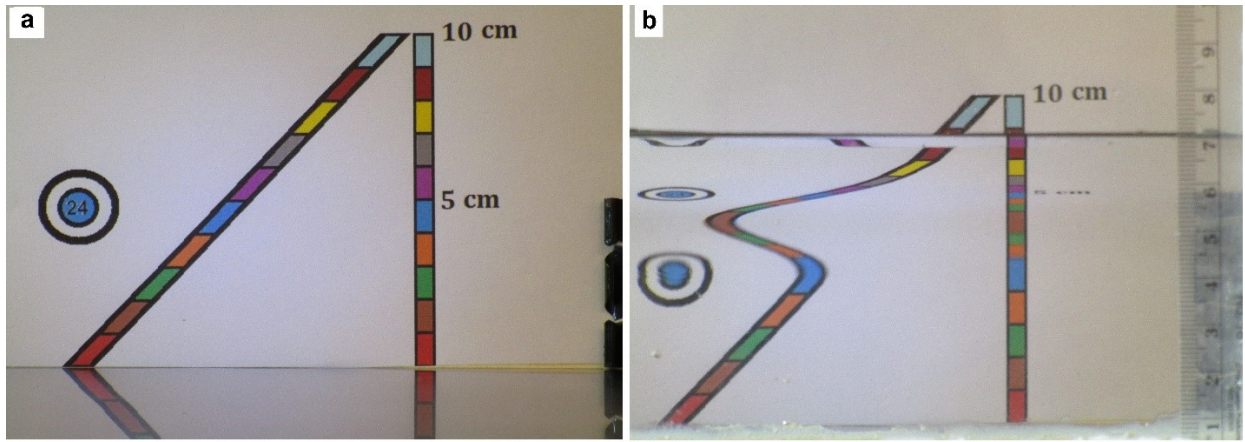

**Fig 12. View through a gradient sucrose solution.** An oblique line graduated with colored centimeter segments (a) is seen through a gradient sucrose solution (b). The region between the first and second bend of the oblique line is an inverted image of the lower portion. Above this region the centimeter segments are erect (non-inverted) and appear compressed. Distance between camera and middle of solution was 80 cm, and between solution and object 30 cm. Room temperature was 20 °C. Internal reflection off the surface of the liquid can be seen at the top. In (a) specular reflection of the lines can be seen on the black epoxy resin table top.

The level of the original sucrose solution (20 g/dL) stood at 5.2 cm above the table top and reached 7.0 cm after layering with distilled water. Fig 12(b) shows three bends of the oblique line separating four regions of solution. The bottom region up to the first bend appears to be uniform sucrose solution. The second bend occurs at approximately where the original solution stood, 5.3 cm. The top 0.5 cm region, after the third bend, appears uniform and would be mostly pure water. The diagram in Fig 13 shows ray crossing for the inverted region and why the upper region appears erect but compressed. It also shows, as Tape [26] did with an aquarium, how light rays from an object can take several paths to produce an image.

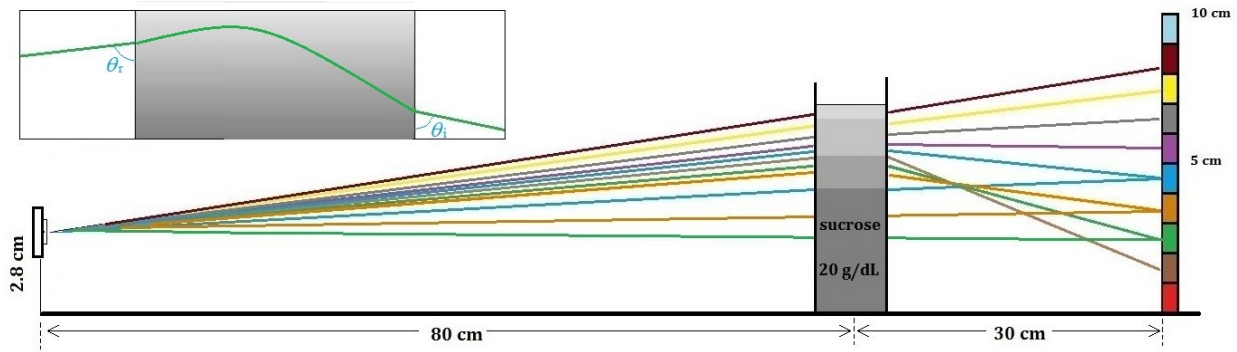

**Fig 13. Ray diagram for Fig 12b.** The shading in the tank shows the four regions of decreasing density. The vertical scale is exaggerated for clarity. The 2- to 4 cm segments appear inverted because their rays cross. The 5- to 8 cm segments appear erect but compressed. The insert shows the path of a single ray changing direction as it reaches the level of the critical angle through the gradient. Note the asymmetry between angles of incidence and angles of reflection.

Solutions with gradient concentrations have been used by Vollmer and Greenler [27] showing dispersion effects with superior mirages. Here, no chromatic aberration was observed with gradient sucrose solutions.

## Specular reflection on rough surfaces

There have been claims that the inferior mirage is due to specular reflection [13–16]. Light impinging on a rough surface at a large angle of incidence exhibits mirror-like reflection. This phenomenon can be seen at incident angles smaller than those observed with desert mirages, and with little or no temperature gradient. Examples of outdoor reflections are shown in Figs 14 and 15a where temperature differences are low or nil. Specular reflection on cloudy days can still be visible when illuminance is low (Fig 15b). However, on sunny days the intensity of diffuse reflection is much greater than that of specular reflection which then becomes obliterated (Fig 15c). This clearly shows that specular reflection on rough surfaces, which is apparently always present, is completely masked by the relatively higher intensity of diffuse reflection on sunny days and thus becomes invisible. Subsequent visits to the road scene in Fig 4a showed illuminance measurements of 5045- and 4936 lx on sunny days, and 2659- and 3422 lx on cloudy days. Thus, on most cloudy days the intensity of diffuse reflection can be sufficiently high to mask specular reflection.

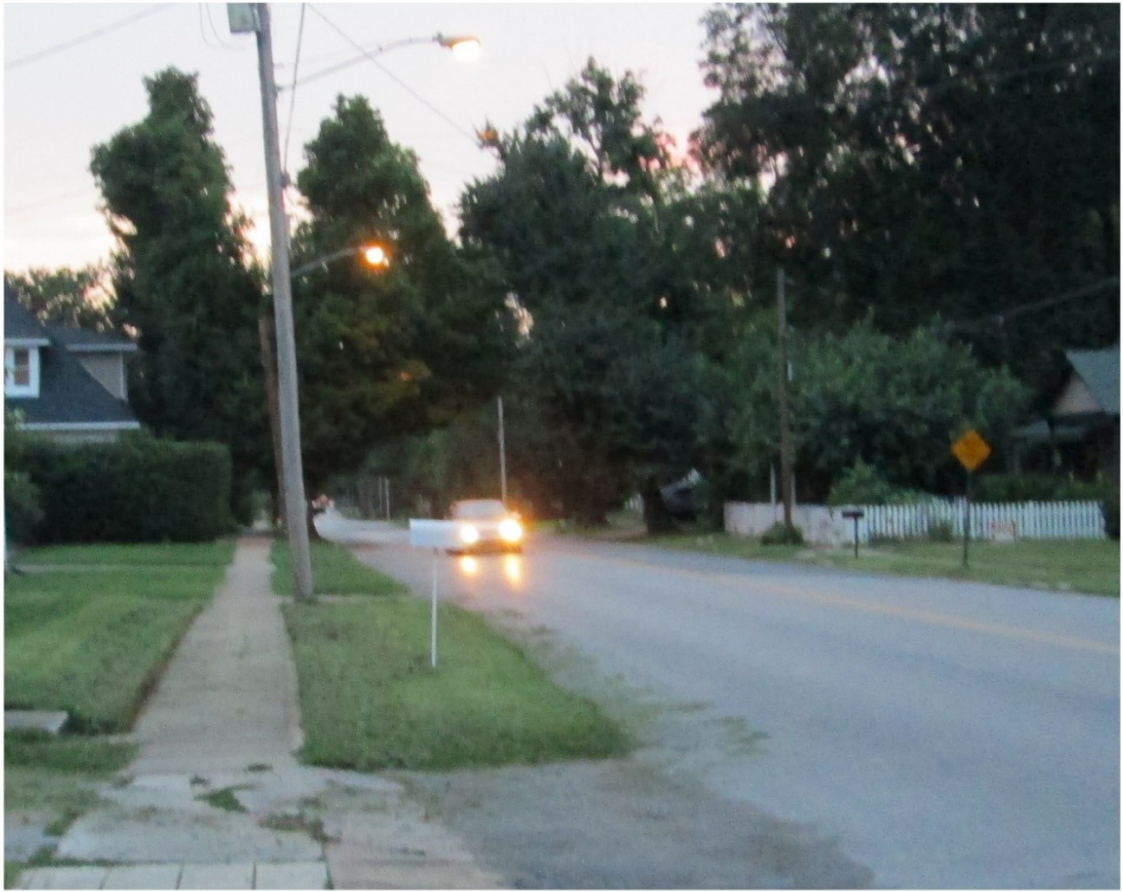

**Fig 14. Specular reflection from rough surfaces.** Distance to reflected object (headlights of car) was approximately 61 m; height 0.6 m; camera height 1.6 m. Angle of incidence was approximately 87.9° determined geometrically (see Fig 2). Temperature difference between ambient and road air was 1.5° C. Photograph taken by the author shortly after sunset.

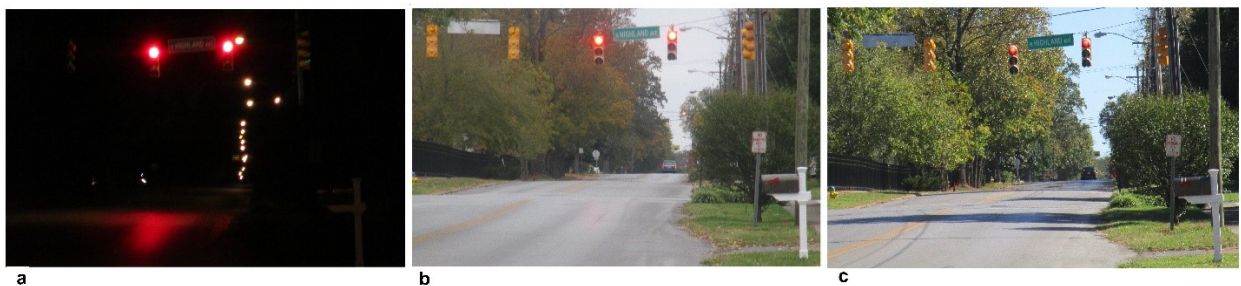

**Fig 15. Specular reflection of traffic lights** at night (a) is still just barely visible on the pavement on a cloudy day (b), but is completely masked on a sunny day (c) by the relatively high intensity of diffuse reflection. Illuminance, measured with an Extech L40 light meter was 0.0 lx (a), 1225 lx (b) and 4477 lx (c). The height of the traffic light, measured by triangulation was 10.9 m, distance 107 m, and camera height 1.6 m. Temperature differences between road and ambient airs were nil in (a), 4 °C in (b) and 7 °C in (c). For the reflection seen in (a) the angle of incidence was approximately 83.3° determined geometrically with Eq 1.

In the laboratory, specular reflection over a black epoxy resin surface shows that the angles of incidence and reflection are virtually the same (Fig 16) as though they are reflected from a mirror laid flat on the table. The asymmetry between angles of incidence and reflection as described by Torrance and Sparrow [28] is not apparent. Note the wide range of viewing angles for specular reflection and the extremely narrow range (here less than 1°) for the refraction type of mirage. (Compare Fig 10b with Fig 16b.) This narrow range is the reason why refraction-type road mirages disappear suddenly as you approach them; with the specular-type of mirage the reflection gradually fades away. Tavassoly *et al.* [15] and Liu and Zhou [16] have given detailed explanations for specular reflection on rough surfaces, but their model does not account for double imaging.

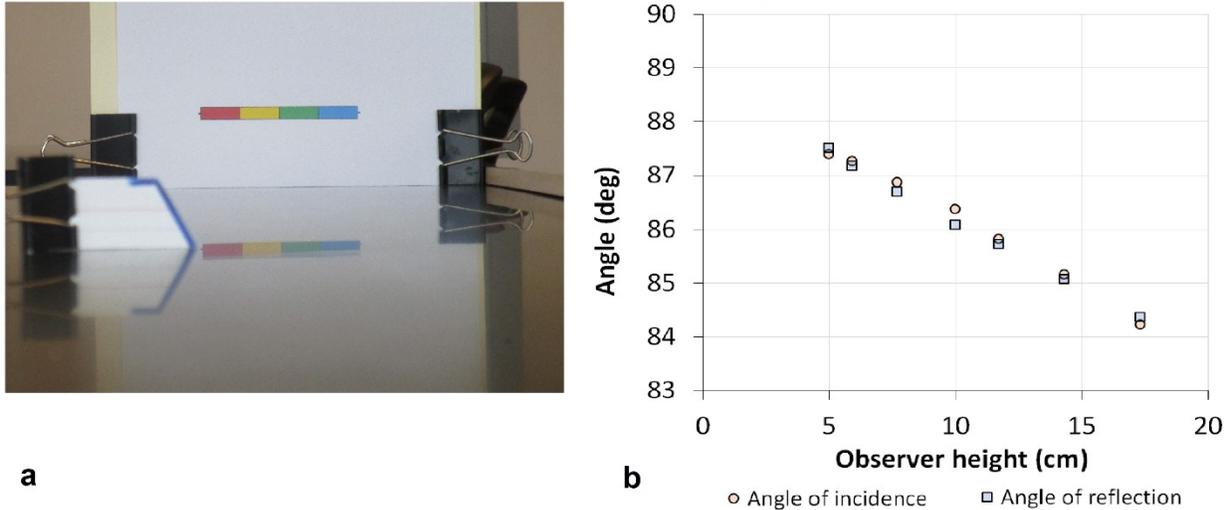

**Fig 16. Specular reflection over black epoxy resin table top.** (a) Movable marker used in locating the position of the reflection. (b) Measured angles of incidence and reflection with respect to observer height.

## Error analysis

In these experiments, the largest errors involved would be in the measurement of temperature and distance. Placing a laboratory thermometer on the pavement or on the sand would measure the temperature of an air layer approximately 0.5 cm thick. This may not show the exact ground temperature and readings would err on the low side, but would remain proportional to the actual temperature as long as the method of measurement is consistent. Distance measured with an odometer involves errors with are difficult to estimate. However, assuming a reasonable error of ± 10%, the angle of incidence (or reflection) in Fig 14 would range from 87.7 to 88.1 degrees. This would still be significantly lower than the angles observed due to refraction (Fig 10).

## Conclusions

Simulation of the desert mirage in the laboratory shows the following.

> A boundary layer of air of only a few millimeters immediately over the heated surface has essentially the same thickness over a wide range of temperatures. Heat transfer in a vertical direction through this layer appears to be in a steady state. The refractive index gradient responsible for reflection is within this layer – not in the gradient above it. It appears to produce both, inverted and erect images of a distant object depending on the observer's distance.

Ray tracing reveals an asymmetry between the angle of incidence and the angle of reflection.

An analogy with sucrose gradient solutions also shows inverted and erect images of a distant object.

Specular reflection over rough surfaces can be described as a mirage, but it is seen at a wider range of angles, it does not require a temperature gradient, and it does not account for double imaging. It is always present, but it can be obliterated by a greater intensity of diffuse reflection. Specular reflection over rough surfaces and reflection due to refraction are two different physical phenomena which are not mutually exclusive.

# Acknowledgements

Thanks are due to professors Chuck Higgins, department of Physics and Astronomy, and to professor William H. Ilsley, department of Chemistry, for valuable discussions.

**Supporting information**

# Temperature profile and double images in the inferior mirage

Nabil W. Wakid

Department of Chemistry, Physical Science Section, Middle Tennessee State University, Murfreesboro, TN 37132

**S1 Table. Effect of temperature difference on the height of the vanishing line.**

| Sand temperature (°C) | $t_g - t_a$ (°C) | measured height from edge **a** (mm) | warmer thickness **b** (mm) | Height of VL from edge (mm) | Observer distance (cm) | Observer height (cm) | Object distance (cm) | IMG |
|---|---|---|---|---|---|---|---|---|
| 40 | 18 | 13.0 | 49 | 7.96 | 320 | 5.9 | 250 | 1482 |
| 50 | 28 | 16.5 | 66 | 7.50 | 320 | 5.9 | 250 | 1483 |
| 58 | 36 | 12.0 | 50 | 7.20 | 320 | 5.9 | 250 | 1484 |
| 61 | 39 | 15.0 | 57 | 7.89 | 320 | 5.9 | 250 | 1487 |
| 64 | 42 | 8.5 | 32 | 7.97 | 320 | 5.9 | 250 | 1490 |
| 36 | 14 | 13.5 | 51 | 7.94 | 320 | 5.9 | 250 | 1493 |
| 45 | 23 | 14.5 | 55 | 7.91 | 320 | 5.9 | 250 | 1494 |
| 50 | 28 | 13.0 | 53 | 7.36 | 320 | 5.9 | 250 | 1495 |
| 52 | 30 | 15.0 | 60 | 7.50 | 320 | 5.9 | 250 | 1496 |
| 54 | 32 | 13.0 | 50 | 7.80 | 320 | 5.9 | 250 | 1497 |
| 39 | 17 | 16.0 | 67 | 7.16 | 320 | 5.65 | 250 | 1593 |
| 45 | 23 | 15.0 | 60 | 7.14 | 320 | 5.65 | 250 | 1595 |

**S2 Table. Temperature profile over heated sand**

| Height (cm) | Temperature (°C) | | |
|---|---|---|---|
| 0 | 32 | 48 | 63 |
| 1 | 25 | 33 | 42 |
| 3 | 21 | 27 | 30 |
| 13 | 20.5 | 23 | 25 |
| 55 | 20 | 20.5 | 21 |
| 89 | 20 | 20 | 20 |

**S3 Table. Effect of observer distance on the angles of incidence and reflection**

| Object dist from reflection (cm) | Object ht from surface (cm) | tan θi | θi (deg) | Observer distance from reflection (cm) | Observer height from surface (cm) | tan θr | θr (deg) | T (°C) | photo IMG |
|---|---|---|---|---|---|---|---|---|---|
| 180 | 0.2 | 900 | 89.94 | 390 | 1.8 | 216.6667 | 89.73556 | 71 | 2120 |
| 210 | 0.1 | 2100 | 89.97 | 360 | 1.8 | 200 | 89.71352 | 71 | 2118 |
| 230 | 0.1 | 2300 | 89.98 | 340 | 1.8 | 188.8889 | 89.69667 | 70 | 2117 |
| 250 | -0.05 | -5000 | 90.01 | 320 | 1.8 | 177.7778 | 89.67771 | 67 | 2115 |
| 285 | -0.1 | -2850 | 90.02 | 285 | 1.8 | 158.3333 | 89.63814 | 63.5 | 2113 |
| 300 | -0.2 | -1500 | 90.04 | 270 | 1.8 | 150 | 89.61803 | 68 | 2111 |

**S4 Table. Effect of observer height on angles of incidence and reflection**

| Observer height from surface (cm) | Object distance from reflection (cm) | Object height from surface (cm) | Observer distance from reflection (cm) | tan $\theta_i$ | $\theta_i$ (deg) | tan $\theta_r$ | $\theta_r$ (deg) | photo IMG |
|---|---|---|---|---|---|---|---|---|
| 1.1 | 285 | 0.5 | 285 | 570 | 89.90 | 259.091 | 89.78 | 2340 |
| 1.3 | 285 | 0.4 | 285 | 712.5 | 89.92 | 219.231 | 89.74 | 2083 |
| 1.8 | 285 | 0.1 | 285 | 2850 | 89.98 | 158.333 | 89.64 | 2113 |
| 2.0 | 285 | -0.1 | 285 | -2850 | 90.02 | 142.5 | 89.60 | 2165 |
| 2.4 | 285 | -0.7 | 285 | -407.14 | 90.14 | 118.75 | 89.52 | 2174 |
| 3.0 | 285 | -0.9 | 285 | -316.67 | 90.18 | 95 | 89.40 | 2182 |

**S5 Table. Average critical angle over heated sand**

| Sand temp (°C) | p (torr) | p (kPa) | temp. wet bulb (°C) | RH at surface (%) | Dew point (°C) | Refractive index at surface | sin $\theta_c$ | $\theta_c$ (deg) |
|---|---|---|---|---|---|---|---|---|
| 22 | 763.5 | 101.8 | 16 | 54 | 12.8 | 1.000270626 | | |
| 67 | 763.5 | 101.8 | - | 6.52 | 12.8 | 1.000234806 | 0.99996419 | 89.52 |
| 71 | 763.5 | 101.8 | - | 5.51 | 12.8 | 1.000232094 | 0.999961479 | 89.50 |

**S6 Table. Specular reflection over black epoxy resin table top**

| object distance from reflection (cm) | object height from surface (cm) | observer distance from reflectn (cm) | observer height from surface (cm) | Tan $\theta_i$ object side | $\theta_i$ (deg) object side | Tan $\theta_r$ observer side | $\theta_r$ (deg) observer side |
|---|---|---|---|---|---|---|---|
| 110 | 5.0 | 115 | 5.0 | 22 | 87.397 | 23 | 87.510 |
| 105 | 5.0 | 120 | 5.9 | 21 | 87.274 | 20.33898 | 87.185 |
| 91.5 | 5.0 | 133.5 | 7.7 | 18.3 | 86.872 | 17.33766 | 86.699 |
| 79 | 5.0 | 146 | 10.0 | 15.8 | 86.379 | 14.6 | 86.082 |
| 68.5 | 5.0 | 156.5 | 11.7 | 13.7 | 85.825 | 13.37607 | 85.724 |
| 59 | 5.0 | 166 | 14.3 | 11.8 | 85.156 | 11.60839 | 85.076 |
| 49.5 | 5.0 | 175.5 | 17.3 | 9.9 | 84.232 | 10.14451 | 84.370 |